\documentclass[aps,prl,twocolumn,superscriptaddress,showpacs,floatfix]{revtex4-1}

\usepackage{graphicx}
\usepackage{dcolumn}
\usepackage{bm}
\usepackage{color}

\begin{document}

\title{Complex magnetism and magnetic field-driven electrical polarization in Co$_3$TeO$_6$}

\author{M. Hudl}\email{matthias.hudl@angstrom.uu.se}
\affiliation{Department of Engineering Sciences, Uppsala University, Box 534, SE-751 21 Uppsala, Sweden}

\author{R. Mathieu}
\affiliation{Department of Engineering Sciences, Uppsala University, Box 534, SE-751 21 Uppsala, Sweden}

\author{S. A. Ivanov}
\affiliation{Department of Engineering Sciences, Uppsala University, Box 534, SE-751 21 Uppsala, Sweden}
\affiliation{Department of Inorganic Materials, Karpov' Institute of Physical Chemistry, Vorontsovo pole, 10 105064, Moscow K-64, Russia}

\author{M. Weil}
\affiliation{Institute for Chemical Technologies and Analytics, Vienna University of Technology, A-1060 Vienna, Austria}

\author{V. Carolus}
\affiliation{HISKP, Universit\"at Bonn, Nussallee 14-16, 53115 Bonn, Germany}

\author{Th. Lottermoser}
\affiliation{HISKP, Universit\"at Bonn, Nussallee 14-16, 53115 Bonn, Germany}
\affiliation{Department of Materials, ETH Zurich, 8093 Zurich, Switzerland}

\author{M. Fiebig}
\affiliation{HISKP, Universit\"at Bonn, Nussallee 14-16, 53115 Bonn, Germany}
\affiliation{Department of Materials, ETH Zurich, 8093 Zurich, Switzerland}

\author{Y. Tokunaga}
\affiliation{Multiferroics Project, ERATO, Japan Science and Technology Agency (JST), Wako 351-0198, Japan}

\author{Y. Taguchi}
\affiliation{Cross-Correlated Materials Research Group (CMRG) and Correlated Electron Research Group (CERG), ASI, RIKEN, 2-1 Hirosawa, Wako 351-0198, Japan} 

\author{Y. Tokura}
\affiliation{Multiferroics Project, ERATO, Japan Science and Technology Agency (JST), Wako 351-0198, Japan}
\affiliation{Cross-Correlated Materials Research Group (CMRG) and Correlated Electron Research Group (CERG), ASI, RIKEN, 2-1 Hirosawa, Wako 351-0198, Japan} 
\affiliation{Department of Applied Physics, University of Tokyo, Tokyo 113-8656, Japan}

\author{P. Nordblad}
\affiliation{Department of Engineering Sciences, Uppsala University, Box 534, SE-751 21 Uppsala, Sweden}

\date{\today}

\begin{abstract}
The magnetic and electrical properties of Co$_3$TeO$_6$ single-crystals with corundum related structure reveal a magnetic-field induced polarization below 21 K. A sharp peak in the specific heat at $\approx$18 K indicates a reconstructive-type first-order phase transition.  From second-harmonic generation (SHG) measurements breaking of inversion symmetry is evident and the point-group symmetry was determined as $m$. The temperature and magnetic-field dependence of the magnetic and electrical polarizations are discussed in the light of the SHG results. 
\end{abstract}


\maketitle


Tailoring materials with coupled magnetic and electrical properties is a challenge in condensed-matter science~\cite{Tokura03,Hur04,Spaldin-fiebig}. For example in a magnetoelectric multiferroic material, the electrical polarization may be controlled by a magnetic field instead of its natural stimuli, an electric field, and vice versa~\cite{Lottermoser04,Gajek07}. In contemporary studies of what is called joint-order-parameter (type II) multiferroics this coupling can be achieved by a variety of mechanisms coupling an improper electric polarization to the magnetic order, e.g., via symmetric or antisymmetric exchange interactions or a spin-dependent d-p hybridization. For example, the antisymmetric Dzyaloshinskii-Moriya interaction can induce an electric polarization according to $P \propto \sum e_{ij} \times (s_i \times s_j)$ where $e_{ij}$ is the vector connecting the two neighboring spins $s_i$ and $s_j$~\cite{Katsura05,Sarma10}. Even though exhibiting a rather small spontaneous electrical polarization the coupling of magnetic and electrical properties in these materials is inherently stronger than that of split-order-parameter (type-I) multiferroics, where magnetic and ferroelectric order emerge independently~\cite{KhomskiiCheong}.
\indent New magnetoelectric multiferroics are thus to be found among materials with magnetic structures dominated by frustration and competing interactions. In this context, it is interesting to investigate physical properties of the sparsely characterized corundum-related $M_3$TeO$_6$ compounds, where M is a first-row transition metal. Ni$_3$TeO$_6$ was found to order antiferromagnetically below 52 K, with a relatively simple magnetic structure~\cite{Zivkovic10}. Mn$_3$TeO$_6$ on the other hand was recently reported to exhibit a complex incommensurate spin structure, consisting of two different magnetic orbits~\cite{Ivanov11}. The magnetic and electrical properties of Co$_3$TeO$_6$ are unknown to date, although its structure shows some interesting features which may strongly influence its magnetic ordering. There are five distinguishable sites for the magnetic Co$^{2+}$ cations and the  $<$Co-O$>$ bond length varies greatly from 1.97 to 2.93 \AA.\\ 
\indent In this communication we report the first results on the magnetic and electrical properties of Co$_3$TeO$_6$ single-crystals. The evolution of the magnetization and polarization of Co$_3$TeO$_6$ with temperature and magnetic field suggests a complex magnetic structure as well as spin rearrangements and spin-flop transitions. The magnetic structure of the compound promotes a magnetic-field induced polarization below 21 K. The origin and nature of the polarization is discussed in the light of optical second-harmonic generation (SHG) data indicating magnetically broken inversion symmetry. \\
\indent High-quality single crystals of Co$_3$TeO$_6$ were grown by chemical transport reactions~\cite{Schaefer63}. A mixture of CoO and TeO$_3$ (obtained by heat treatment of Co(NO$_3$)$_2\cdot$6H$_2$O and H$_6$TeO$_6$ respectively) in the stoichiometric ratio 3:1 was thoroughly ground and was, together with 10 mg PtCl$_2$, placed in a silica ampoule which was evacuated, sealed, and heated in a temperature gradient 1098 $\rightarrow$ 1028 K.  After 5 days the transport reaction was completed and dark blue to black crystals of Co$_3$TeO$_6$, mostly with a prismatic pinacoidal form and edge-lengths up to 5 mm, had formed in the colder part of the ampoule. X-ray powder diffraction patterns were recorded on crushed single crystals using a Bruker D-8 diffractometer with CuK$\alpha$ radiation, and analyzed with Rietveld techniques using \textsc{Fullprof} software.  \newpage

\begin{figure}[htb]
 \includegraphics[width=0.40\textwidth]{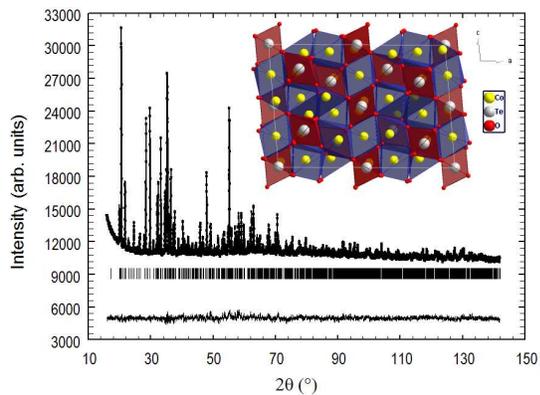}
	\caption{(Color Online) X-ray diffraction pattern and Rietveld refinement for Co$_3$TeO$_6$ at room temperature (Rietveld residuals: $R_p$ = 2.07\%, $R_{wp}$ = 3.49\%, $R_b$ = 6.42\%. See Ref.  [\onlinecite{ivanov111}] for details; the Rietveld analysis confirmed the absence of impurity phases). Inset presents an polyhedral representation of the structure.}
	\label{CTO1}
\end{figure} 

Magnetization measurements up to 20 Oe were performed using a Quantum Design MPMS XL SQUID magnetometer following zero-field-cooled (ZFC) and field-cooled (FC) protocols, while magnetization measurements above 20 Oe (using a VSM),  heat capacity (using a relaxation method) as well as pyroelectric measurements were performed on a Quantum Design PPMS. In order to check the magnetoelectric response; thin plates of Co$_3$TeO$_6$ crystals, oriented perpendicular to the $a$-axis~\cite{note} and the $c$-axis respectively, were prepared. In the experiments, these were poled in a $\approx$ 500 V/mm electric field and cooled from 50 K in the PPMS. 
The electrical polarization $P$ was obtained from integrating the pyroelectric current recorded using an electrometer (Keithley 6517A). For nonlinear optical experiments~\cite{Fiebig111} crystals were polished to about 50~$\mu$m thickness with orientations perpendicular to the crystallographic $a$, $b$ and $c$ axis, respectively. The samples were mounted in a He-vapor-operated cryostat and excited by light pulses (5 ns, 10-40 Hz, 2.3-2.9 eV) generated with an optical parametric oscillator pumped by a Nd:YAG laser. Signals were detected by a liquid-nitrogen-cooled digital camera.\\
\indent The X-ray diffraction pattern and associated Rietveld refinement at 296 K is shown in Fig.~\ref{CTO1}. Co$_3$TeO$_6$ adopts a monoclinic structure (space group $C2/c$), with lattice parameters $a$ = 14.8014(3)\AA, $b$ = 8.8379(2)\AA, $c$ = 10.3421(3)\AA, and monoclinic angle $\beta$ = 94.83(1). These values are in good agreement with those of the previous single crystal study~\cite{Becker06}. 
There are five distinguishable Co$^{2+}$ sites, out of which three are octahedrally, one is square-pyramidally and one is tetrahedrally coordinated. The two distinct Te$^{6+}$ sites have an octahedral coordination~\cite{Becker06}. \\
 
\begin{figure}[htb]
	\includegraphics[width=0.40\textwidth]{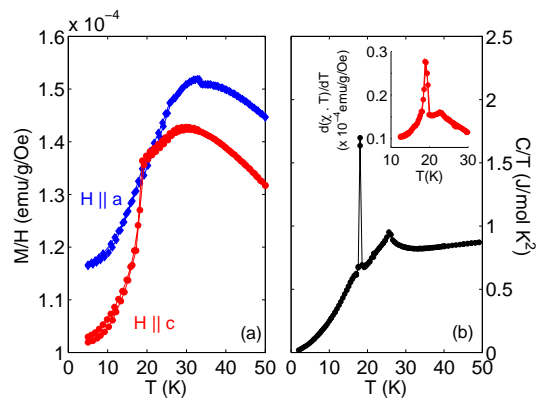}
	\caption{(Color online) Temperature dependence of (a) the zero-field
 	cooled (ZFC) and field-cooled (FC) magnetization in 20 Oe magnetic field for different orientations, and (b) heat capacity, plotted as $C/T$. No significant differences were observed in heat capacity measurements recorded on cooling and heating~\cite{Nirmala11}. The inset 	shows the temperature dependence of $d \chi T/dT$ ($\chi=M/H$) calculated from the magnetization data with $H||c$.}
	\label{CTO2}
\end{figure}

\indent Low-magnetic field (20 Oe) ZFC and FC magnetization data of Co$_3$TeO$_6$ for different orientations of the applied magnetic field $H$ are shown in Fig.~\ref{CTO2}(a). The magnetization recorded with $H$ applied along the $a$-axis and the $c$-axis exhibits a maximum close to 30 K suggesting antiferromagnetic ordering. In both cases a marginal irreversibility between the ZFC/FC curves is observed at low temperature. For the magnetization recorded with  $H||a$ a small but distinct step of unknown nature is observed in both the ZFC and FC curves just above 30 K. With $H||c$, a sharp drop in the magnetization is observed at T* $\sim$ 18 K.  A Curie-Weiss analysis of the high-temperature magnetization data was performed, yielding a Weiss constant of $\theta$ = $-$54 K and an effective Bohr magneton number of $p$ = 4.69. The $\theta$ and $p$ values confirm an (antiferro)magnetic interaction and covalency effects, respectively~\cite{Mathieu11}. The heat-capacity data shown in Fig.~\ref{CTO2}(b) confirms that a long-range (antiferromagnetic) order is established near 26 K. Interestingly, a sharp peak is observed at 18 K, i.e. in the vicinity of T*. Such a $\delta$-peak-like anomaly in the heat capacity could be an indication for a reconstructive first-order phase transition~\cite{Toledano}. Both features at 26 K and 18 K are evident in the temperature derivative of the magnetic susceptibility ($\chi=M/H$) times temperature ($T$), $d(\chi \cdot T)/dT$, as seen in inset of Fig.~\ref{CTO2}(b). ($(d(\chi \cdot T)/dT$) should be proportional to the magnetic specific heat for an antiferromagnet~\cite{Fisher62}). Neutron powder diffraction experiments on crushed single crystals suggest a magnetic structure involving several propagation vectors~\cite{ivanov111}. \\

\begin{figure}[htb]
   \includegraphics[width=0.40\textwidth]{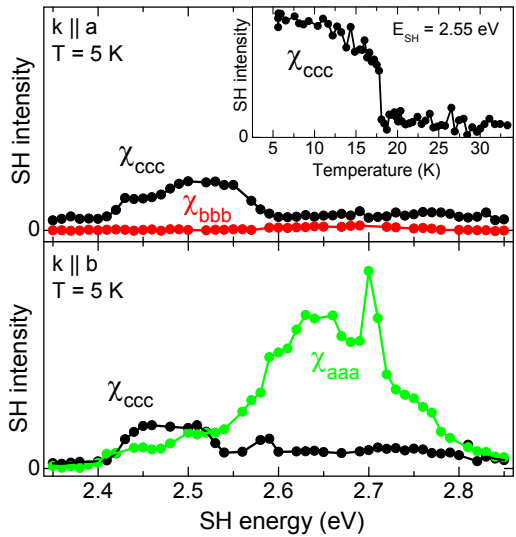}
	\caption{(Color online) Second harmonic generation at 5 K for $k$ $||$ $a$ (upper panel) and $k$ $||$ $b$ (lower panel). $k$ represents the wave vector of the incident light and $\chi_{aaa}$ and $\chi_{ccc}$ are SHG susceptibility tensor components; the red line depicts $\chi_{bbb}$ which does not contribute to the SHG signal. The inset shows the temperature dependence of $\chi_{ccc}$.} 
	\label{CTO3}
\end{figure}

\indent The symmetry properties of long-range ordered and, in particular, non-centrosymmetric compounds can be probed by SHG. In the electric-dipole approximation SHG can be described by $P_{i}(2\omega)=\epsilon_{0}\, \chi_{ijk}\, E_{j}(\omega)\, E_{k}(\omega)$, where $\vec{E}(\omega)$ and $\vec{P}(2\omega)$ represent the incident light at frequency $\omega$ and the polarization induced at frequency $2\omega$~\cite{Fiebig111}. The set of nonzero tensor components $\chi_{ijk}$ is determined by the symmetry of the material. In the case of inversion symmetry we have $\chi_{ijk}\equiv0$. Therefore one would not expect any SHG signal for Co$_3$TeO$_6$ in the non-ferroic phase above $T_N$ with the space group $C2/c$ (point group $2/m$). A broken inversion symmetry in the low temperature ferroic phase would reduce the point symmetry to $m$, 2 or 1, respectively~\cite{Birss} and induce non-zero components according to: m ($\chi_{aaa}$, $\chi_{ccc}$), 2 ($\chi_{bbb}$), 1 ($\chi_{aaa}$, $\chi_{bbb}$, $\chi_{ccc}$). As shown in Fig.~\ref{CTO3} there are $\chi_{aaa}$ and $\chi_{ccc}$ contributions to the SHG in the $a$ and $b$ cut samples in the low temperature phase, but no signal for the $\chi_{bbb}$ component. This reveals $m$ as the point symmetry for this phase. For point-group symmetry $m$ a polar axis in the $ac$-plane perpendicular to the monoclinic $b$-axis exist~\cite{Birss}. This means a possible ferroelectric polarization would be oriented parallel to this polar axis. Furthermore, in the inset of Fig.~\ref{CTO3}, the ferroic transition at 18 K is evidenced from the temperature dependence of $\chi_{ccc}$. Notwithstanding that our SHG data indicate magnetically broken inversion symmetry a possible absence of the spontaneous polarization needs to be considered. 

\begin{figure}[htb]
 	\includegraphics[width=0.40\textwidth]{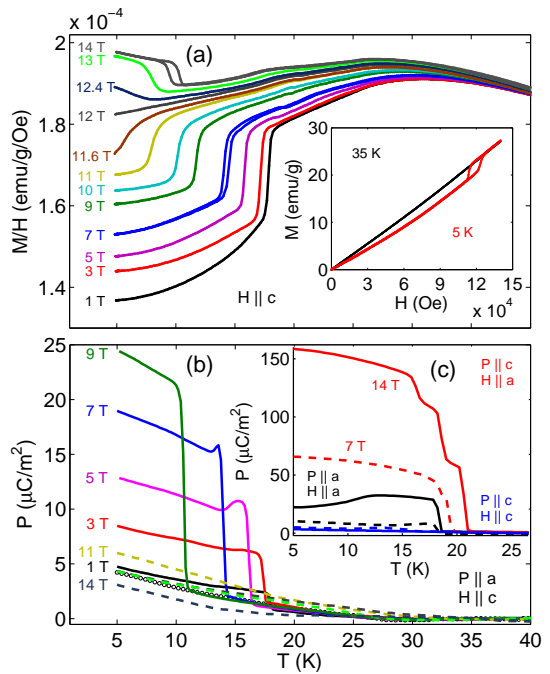}
	\caption{(Color online) (a) Field-cooled magnetization and (b) electric polarization as a function of temperature in several magnetic fields up to 14 T applied along the $c$-axis. The magnetization curves for 7 T and 14 T field are shown for both cooling and heating sweeps. Inset in (a) shows magnetization as a function of magnetic field at 5 K and 35 K. In (c) the temperature dependence of the polarization for applied magnetic fields of 7 T and 14 T for different orientations of the applied magnetic field and measured polarization is presented.} 
	\label{CTO4}
\end{figure}

\begin{figure}[htb]
 \includegraphics[width=0.40\textwidth]{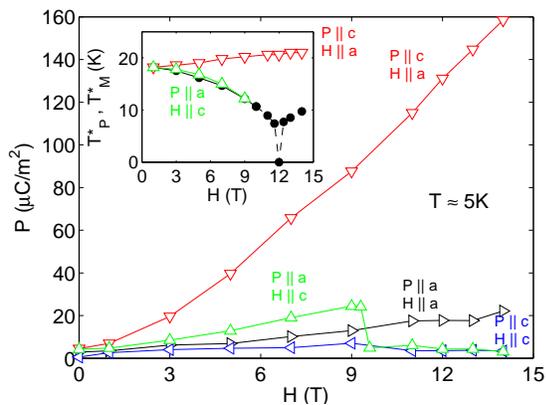}
	\caption{(Color online) Low-temperature polarization as a function of magnetic field for the different $P$ and $H$ orientations considered. The inset shows the variation of the temperature onset of polarization $T_P^*$ for ($P||c$, $H||a$) (\textcolor{red}{$\bigtriangledown$}) and ($P||a$, $H||c$) (\textcolor{green}{$\bigtriangleup$}), as well as the temperature $T_M^*$ (\textcolor{black}{$\bullet$}) below which a drop/step is observed in the magnetization recorded with $H||c$.} 
	\label{CTO5}
\end{figure}

Magnetic and electrical properties in high-magnetic fields up to 14 Tesla are depicted in Fig.~\ref{CTO4}. The FC cooling magnetization measurements in $c$-axis orientation reveal that the drop in magnetization (observed at 18 K in low-fields) is shifted towards lower temperatures with increasing applied magnetic fields. For magnetic fields larger than 11 T an upward step is observed in the magnetization data instead of a drop. The rearrangement of the magnetization at low temperatures can also be perceived in magnetization vs. magnetic field measurements, such as the ones shown in the inset of Fig.~\ref{CTO4}(a). \\
\indent At 5 K, a possible spontaneous polarization of the order 5 $\mu$C/m$^2$ (close to the resolution limit of our experimental set-up) was measured, see Fig.~\ref{CTO4}(b) (open data markers).When applying magnetic fields relatively large polarization values are induced below 21 K. As can be seen in Fig.~\ref{CTO4}(b), the magnetic-field induced polarization ($P||a$, $H||c$) reaches a value of about 25 $\mu$C/m$^2$ at 5 K and 9 T. When the magnetic field is increased beyond 9.3 T the electrical polarization gets rapidly suppressed. The temperature dependence of the magnetic-field induced polarization at 7 T and 14 T field recorded for different orientations of the applied magnetic field and measured polarization is depicted in Fig.~\ref{CTO4}(c). For orientation ($P||c$, $H||a$) the polarization steadily increases with increasing magnetic field, and reaches about 160 $\mu$C/m$^2$ in 14 T. The electric polarization is strongly interconnected with magnetic behavior, as illustrated in the inset of Fig.~\ref{CTO5}. The onset of the field-induced polarization ($P||a$, $H||c$) occurs at the same temperatures as the drop (or step) in the magnetization $T_M^*$. Furthermore, the temperature onset of the induced polarization $T_P^*$ has an orientational dependence on the magnetic field, as seen in the same inset. For $H||a$ the temperature onset of polarization increases with increasing magnetic field whereas for $H||c$ the polarization onset decreases. \\
\indent Two characteristic temperatures were observed in the low-field magnetization, polarization, heat capacity and neutron data~\cite{ivanov111}, T$_N$ ($\sim$26 K) and T* ($\sim$18 K). At T$_N$, the spins may form a sinusoidal antiferromagnetic arrangement directed along the $c$-axis which turns into a spin spiral arrangement below T*. According to SHG, the point symmetry of the low-temperature phase is $m$, suggesting that the orientation of a possible electrical polarization would be in the "$ac$"-plane~\cite{Birss}. In view of the present data it is difficult to identify the mechanism for the magnetic-field induced polarization in Co$_3$TeO$_6$. However, considering several characteristic features, such as: an incommensurate magnetic structure, a relatively large field-orientation dependence of the induced polarization, and a relatively small value of induced polarization; symmetric exchange striction should not be relevant. On the other hand, from the $H$-dependence of $P$ shown in Fig.~\ref{CTO5}, the low-temperature polarization both along $a$ and $c$ increases as $H||a$ is increased. In case of a spiral magnetic structure, it may be difficult to attribute these observations to an antisymmetric exchange mechanism alone, while a gradual rotation of the spiral plane and the subsequent increase of the cycloidal components may explain the $H$-dependence of the polarization in the case of $P||c$. \\
\indent To conclude, we have investigated the magnetic, electrical, specific heat, and non-linear optical properties of Co$_3$TeO$_6$ single-crystals. Specific heat data reveal (antiferro)magnetic ordering at $\sim$26 K and a first-order-like phase transition at $\sim$18 K, both likewise observed in the magnetic measurements. The magnetic structure brings forth a magnetic-field induced electrical polarization. For magnetic fields $H||c$ the induced polarization is suppressed above 9.3 T whereas for $H||a$ a significant enhancement of $P||c$ up to 160 $\mu$C/m$^2$ is observed in 14 T. Our results suggest a strong coupling for the magnetic and electrical structure of Co$_3$TeO$_6$ below 21 K. We therefore believe that corundum-related compounds such as Co$_3$TeO$_6$ are a new interesting class of magnetoelectric multiferroic systems. \\

\begin{acknowledgments}
We thank P. Tol\'{e}dano for discussions. We acknowledge and thank the Swedish Research Council (VR), the G\"oran Gustafsson Foundation, Anna-Maria Lundin foundation, Hans Werth\'en Fonden, the FIRST Program by JSPS, SFB608 of the DFG, and the Russian Foundation for Basic Research for financial support. 
\end{acknowledgments}

\bibliographystyle{apsrev4-1}

\begin{thebibliography}{20}%
\makeatletter
\providecommand \@ifxundefined [1]{%
 \@ifx{#1\undefined}
}%
\providecommand \@ifnum [1]{%
 \ifnum #1\expandafter \@firstoftwo
 \else \expandafter \@secondoftwo
 \fi
}%
\providecommand \@ifx [1]{%
 \ifx #1\expandafter \@firstoftwo
 \else \expandafter \@secondoftwo
 \fi
}%
\providecommand \natexlab [1]{#1}%
\providecommand \enquote  [1]{``#1''}%
\providecommand \bibnamefont  [1]{#1}%
\providecommand \bibfnamefont [1]{#1}%
\providecommand \citenamefont [1]{#1}%
\providecommand \href@noop [0]{\@secondoftwo}%
\providecommand \href [0]{\begingroup \@sanitize@url \@href}%
\providecommand \@href[1]{\@@startlink{#1}\@@href}%
\providecommand \@@href[1]{\endgroup#1\@@endlink}%
\providecommand \@sanitize@url [0]{\catcode `\\12\catcode `\$12\catcode
  `\&12\catcode `\#12\catcode `\^12\catcode `\_12\catcode `\%12\relax}%
\providecommand \@@startlink[1]{}%
\providecommand \@@endlink[0]{}%
\providecommand \url  [0]{\begingroup\@sanitize@url \@url }%
\providecommand \@url [1]{\endgroup\@href {#1}{\urlprefix }}%
\providecommand \urlprefix  [0]{URL }%
\providecommand \Eprint [0]{\href }%
\providecommand \doibase [0]{http://dx.doi.org/}%
\providecommand \selectlanguage [0]{\@gobble}%
\providecommand \bibinfo  [0]{\@secondoftwo}%
\providecommand \bibfield  [0]{\@secondoftwo}%
\providecommand \translation [1]{[#1]}%
\providecommand \BibitemOpen [0]{}%
\providecommand \bibitemStop [0]{}%
\providecommand \bibitemNoStop [0]{.\EOS\space}%
\providecommand \EOS [0]{\spacefactor3000\relax}%
\providecommand \BibitemShut  [1]{\csname bibitem#1\endcsname}%
\let\auto@bib@innerbib\@empty
\bibitem [{\citenamefont {Kimura}\ \emph {et~al.}(2003)\citenamefont {Kimura},
  \citenamefont {Goto}, \citenamefont {Shintani}, \citenamefont {Ishizaka},
  \citenamefont {Arima},\ and\ \citenamefont {Tokura}}]{Tokura03}%
  \BibitemOpen
  \bibfield  {author} {\bibinfo {author} {\bibfnamefont {T.}~\bibnamefont
  {Kimura}}, \bibinfo {author} {\bibfnamefont {T.}~\bibnamefont {Goto}},
  \bibinfo {author} {\bibfnamefont {H.}~\bibnamefont {Shintani}}, \bibinfo
  {author} {\bibfnamefont {K.}~\bibnamefont {Ishizaka}}, \bibinfo {author}
  {\bibfnamefont {T.}~\bibnamefont {Arima}}, \ and\ \bibinfo {author}
  {\bibfnamefont {Y.}~\bibnamefont {Tokura}},\ }\href@noop {} {\bibfield
  {journal} {\bibinfo  {journal} {Nature}\ }\textbf {\bibinfo {volume} {426}},\
  \bibinfo {pages} {55} (\bibinfo {year} {2003})}\BibitemShut {NoStop}%
\bibitem [{\citenamefont {Hur}\ \emph {et~al.}(2004)\citenamefont {Hur},
  \citenamefont {Park}, \citenamefont {Sharma}, \citenamefont {Ahn},
  \citenamefont {Guha},\ and\ \citenamefont {Cheong}}]{Hur04}%
  \BibitemOpen
  \bibfield  {author} {\bibinfo {author} {\bibfnamefont {H.}~\bibnamefont
  {Hur}}, \bibinfo {author} {\bibfnamefont {S.}~\bibnamefont {Park}}, \bibinfo
  {author} {\bibfnamefont {P.~A.}\ \bibnamefont {Sharma}}, \bibinfo {author}
  {\bibfnamefont {J.~S.}\ \bibnamefont {Ahn}}, \bibinfo {author} {\bibfnamefont
  {S.}~\bibnamefont {Guha}}, \ and\ \bibinfo {author} {\bibfnamefont {S.-W.}\
  \bibnamefont {Cheong}},\ }\href@noop {} {\bibfield  {journal} {\bibinfo
  {journal} {Nature}\ }\textbf {\bibinfo {volume} {429}},\ \bibinfo {pages}
  {392} (\bibinfo {year} {2004})}\BibitemShut {NoStop}%
\bibitem [{\citenamefont {Spaldin}\ and\ \citenamefont
  {Fiebig}()}]{Spaldin-fiebig}%
  \BibitemOpen
  \bibfield  {author} {\bibinfo {author} {\bibfnamefont {N.~A.}\ \bibnamefont
  {Spaldin}}\ and\ \bibinfo {author} {\bibfnamefont {M.}~\bibnamefont
  {Fiebig}},\ }\href@noop {} {\bibinfo  {journal} {Science \textbf{309}, 391
  (2005); N. A. Hill, J. Phys. Chem. B \textbf{104}, 6694 (2000); M. Fiebig, J.
  Phys. D \textbf{38}, R123 (2005)}\ }\BibitemShut {NoStop}%
\bibitem [{\citenamefont {Lottermoser}\ \emph {et~al.}(2004)\citenamefont
  {Lottermoser}, \citenamefont {Lonkai}, \citenamefont {Amann}, \citenamefont
  {Hohlwein}, \citenamefont {Ihringer},\ and\ \citenamefont
  {Fiebig}}]{Lottermoser04}%
  \BibitemOpen
\bibfield  {journal} {  }\bibfield  {author} {\bibinfo {author} {\bibfnamefont
  {Th.}~\bibnamefont {Lottermoser}}, \bibinfo {author} {\bibfnamefont
  {T.}~\bibnamefont {Lonkai}}, \bibinfo {author} {\bibfnamefont
  {U.}~\bibnamefont {Amann}}, \bibinfo {author} {\bibfnamefont
  {D.}~\bibnamefont {Hohlwein}}, \bibinfo {author} {\bibfnamefont
  {J.}~\bibnamefont {Ihringer}}, \ and\ \bibinfo {author} {\bibfnamefont
  {M.}~\bibnamefont {Fiebig}},\ }\href@noop {} {\bibfield  {journal} {\bibinfo
  {journal} {Nature}\ }\textbf {\bibinfo {volume} {430}},\ \bibinfo {pages}
  {541} (\bibinfo {year} {2004})}\BibitemShut {NoStop}%
\bibitem [{\citenamefont {Gajek}\ \emph {et~al.}(2007)\citenamefont {Gajek},
  \citenamefont {Bibes}, \citenamefont {Fusil}, \citenamefont {Bouzehouane},
  \citenamefont {Fontcuberta}, \citenamefont {Barthelemy},\ and\ \citenamefont
  {Fert}}]{Gajek07}%
  \BibitemOpen
  \bibfield  {author} {\bibinfo {author} {\bibfnamefont {M.}~\bibnamefont
  {Gajek}}, \bibinfo {author} {\bibfnamefont {M.}~\bibnamefont {Bibes}},
  \bibinfo {author} {\bibfnamefont {S.}~\bibnamefont {Fusil}}, \bibinfo
  {author} {\bibfnamefont {K.}~\bibnamefont {Bouzehouane}}, \bibinfo {author}
  {\bibfnamefont {J.}~\bibnamefont {Fontcuberta}}, \bibinfo {author}
  {\bibfnamefont {A.}~\bibnamefont {Barthelemy}}, \ and\ \bibinfo {author}
  {\bibfnamefont {A.}~\bibnamefont {Fert}},\ }\href@noop {} {\bibfield
  {journal} {\bibinfo  {journal} {Nature Mater.}\ }\textbf {\bibinfo {volume}
  {6}},\ \bibinfo {pages} {296} (\bibinfo {year} {2007})}\BibitemShut {NoStop}%
\bibitem [{\citenamefont {Katsura}\ \emph {et~al.}(2005)\citenamefont
  {Katsura}, \citenamefont {Nagaosa},\ and\ \citenamefont
  {Balatsky}}]{Katsura05}%
  \BibitemOpen
  \bibfield  {author} {\bibinfo {author} {\bibfnamefont {H.}~\bibnamefont
  {Katsura}}, \bibinfo {author} {\bibfnamefont {N.}~\bibnamefont {Nagaosa}}, \
  and\ \bibinfo {author} {\bibfnamefont {A.~V.}\ \bibnamefont {Balatsky}},\
  }\href@noop {} {\bibfield  {journal} {\bibinfo  {journal} {Phys. Rev. Lett.}\
  }\textbf {\bibinfo {volume} {95}},\ \bibinfo {pages} {057205} (\bibinfo
  {year} {2005})}\BibitemShut {NoStop}%
\bibitem [{\citenamefont {Shanavas}\ \emph {et~al.}(2010)\citenamefont
  {Shanavas}, \citenamefont {Choudhury}, \citenamefont {Dasgupta},
  \citenamefont {Sharma},\ and\ \citenamefont {Sarma}}]{Sarma10}%
  \BibitemOpen
  \bibfield  {author} {\bibinfo {author} {\bibfnamefont {K.~V.}\ \bibnamefont
  {Shanavas}}, \bibinfo {author} {\bibfnamefont {D.}~\bibnamefont {Choudhury}},
  \bibinfo {author} {\bibfnamefont {I.}~\bibnamefont {Dasgupta}}, \bibinfo
  {author} {\bibfnamefont {S.~M.}\ \bibnamefont {Sharma}}, \ and\ \bibinfo
  {author} {\bibfnamefont {D.~D.}\ \bibnamefont {Sarma}},\ }\href {\doibase
  10.1103/PhysRevB.81.212406} {\bibfield  {journal} {\bibinfo  {journal} {Phys.
  Rev. B}\ }\textbf {\bibinfo {volume} {81}},\ \bibinfo {pages} {212406}
  (\bibinfo {year} {2010})}\BibitemShut {NoStop}%
\bibitem [{\citenamefont {Khomskii}()}]{KhomskiiCheong}%
  \BibitemOpen
  \bibfield  {author} {\bibinfo {author} {\bibfnamefont {D.}~\bibnamefont
  {Khomskii}},\ }\href@noop {} {\bibinfo  {journal} {Physics \textbf{2}, 20
  (2009); S.-W. Cheong and M. Mostovoy, Nature Mater. \textbf{6}, 13 (2007)}\
  }\BibitemShut {NoStop}%
\bibitem [{\citenamefont {Zivkovic}\ \emph {et~al.}(2010)\citenamefont
  {Zivkovic}, \citenamefont {Prsa}, \citenamefont {Zaharko},\ and\
  \citenamefont {Berger}}]{Zivkovic10}%
  \BibitemOpen
\bibfield  {journal} {  }\bibfield  {author} {\bibinfo {author} {\bibfnamefont
  {I.}~\bibnamefont {Zivkovic}}, \bibinfo {author} {\bibfnamefont
  {K.}~\bibnamefont {Prsa}}, \bibinfo {author} {\bibfnamefont {O.}~\bibnamefont
  {Zaharko}}, \ and\ \bibinfo {author} {\bibfnamefont {H.}~\bibnamefont
  {Berger}},\ }\href@noop {} {\bibfield  {journal} {\bibinfo  {journal} {J.
  Phys.: Condens. Matter}\ }\textbf {\bibinfo {volume} {22}},\ \bibinfo {pages}
  {056002} (\bibinfo {year} {2010})}\BibitemShut {NoStop}%
\bibitem [{\citenamefont {Ivanov}\ \emph {et~al.}()\citenamefont {Ivanov},
  \citenamefont {Nordblad}, \citenamefont {Mathieu}, \citenamefont {Tellgren},
  \citenamefont {Ritter}, \citenamefont {Golubko}, \citenamefont {Politova},\
  and\ \citenamefont {Weil}}]{Ivanov11}%
  \BibitemOpen
  \bibfield  {author} {\bibinfo {author} {\bibfnamefont {S.~A.}\ \bibnamefont
  {Ivanov}}, \bibinfo {author} {\bibfnamefont {P.}~\bibnamefont {Nordblad}},
  \bibinfo {author} {\bibfnamefont {R.}~\bibnamefont {Mathieu}}, \bibinfo
  {author} {\bibfnamefont {R.}~\bibnamefont {Tellgren}}, \bibinfo {author}
  {\bibfnamefont {C.}~\bibnamefont {Ritter}}, \bibinfo {author} {\bibfnamefont
  {N.}~\bibnamefont {Golubko}}, \bibinfo {author} {\bibfnamefont {E.~D.}\
  \bibnamefont {Politova}}, \ and\ \bibinfo {author} {\bibfnamefont
  {M.}~\bibnamefont {Weil}},\ }\href@noop {} {\bibinfo  {journal} {to appear in
  Mater. Res. Bull. (2011); cond-mat/1104.5560}\ }\BibitemShut {NoStop}%
\bibitem [{\citenamefont {Sch\"afer}(1963)}]{Schaefer63}%
  \BibitemOpen
\bibfield  {journal} {  }\bibfield  {author} {\bibinfo {author} {\bibfnamefont
  {H.}~\bibnamefont {Sch\"afer}},\ }\href@noop {} {\emph {\bibinfo {title}
  {{Chemical Transport Reactions}}}}\ (\bibinfo  {publisher} {Academic Press -
  New York},\ \bibinfo {year} {1963})\BibitemShut {NoStop}%
\bibitem [{\citenamefont {Ivanov~et al.}()}]{ivanov111}%
  \BibitemOpen
  \bibfield  {author} {\bibinfo {author} {\bibfnamefont {S.~A.}\ \bibnamefont
  {Ivanov~et al.}},\ }\href@noop {} {\bibinfo  {journal} {unpublished}\
  }\BibitemShut {NoStop}%
\bibitem [{not()}]{note}%
  \BibitemOpen
\bibfield  {journal} {  }\href@noop {} {\bibinfo  {journal} {The $a$-axis
  sample for the electrical polarization measurements may involve a $b$-axis
  component such as [310]}\ }\BibitemShut {NoStop}%
\bibitem [{\citenamefont {Fiebig}\ \emph {et~al.}(2005)\citenamefont {Fiebig},
  \citenamefont {Pavlov},\ and\ \citenamefont {Pisarev}}]{Fiebig111}%
  \BibitemOpen
\bibfield  {journal} {  }\bibfield  {author} {\bibinfo {author} {\bibfnamefont
  {M.}~\bibnamefont {Fiebig}}, \bibinfo {author} {\bibfnamefont {V.~V.}\
  \bibnamefont {Pavlov}}, \ and\ \bibinfo {author} {\bibfnamefont {R.~V.}\
  \bibnamefont {Pisarev}},\ }\href@noop {} {\bibfield  {journal} {\bibinfo
  {journal} {J. Opt. Soc. Am. B}\ }\textbf {\bibinfo {volume} {22}},\ \bibinfo
  {pages} {96} (\bibinfo {year} {2005})}\BibitemShut {NoStop}%
\bibitem [{\citenamefont {Becker}\ \emph {et~al.}(2006)\citenamefont {Becker},
  \citenamefont {Johnsson},\ and\ \citenamefont {Berger}}]{Becker06}%
  \BibitemOpen
  \bibfield  {author} {\bibinfo {author} {\bibfnamefont {R.}~\bibnamefont
  {Becker}}, \bibinfo {author} {\bibfnamefont {M.}~\bibnamefont {Johnsson}}, \
  and\ \bibinfo {author} {\bibfnamefont {H.}~\bibnamefont {Berger}},\
  }\href@noop {} {\bibfield  {journal} {\bibinfo  {journal} {Acta Cryst.}\
  }\textbf {\bibinfo {volume} {C62}},\ \bibinfo {pages} {i67} (\bibinfo {year}
  {2006})}\BibitemShut {NoStop}%
\bibitem [{\citenamefont {Nirmala}\ \emph {et~al.}(2011)\citenamefont
  {Nirmala}, \citenamefont {Paudyal}, \citenamefont {Pecharsky}, \citenamefont
  {Gschneidner~Jr.},\ and\ \citenamefont {Nigam}}]{Nirmala11}%
  \BibitemOpen
  \bibfield  {author} {\bibinfo {author} {\bibfnamefont {R.}~\bibnamefont
  {Nirmala}}, \bibinfo {author} {\bibfnamefont {D.}~\bibnamefont {Paudyal}},
  \bibinfo {author} {\bibfnamefont {V.~K.}\ \bibnamefont {Pecharsky}}, \bibinfo
  {author} {\bibfnamefont {K.~A.}\ \bibnamefont {Gschneidner~Jr.}}, \ and\
  \bibinfo {author} {\bibfnamefont {A.~K.}\ \bibnamefont {Nigam}},\ }\href@noop
  {} {\bibfield  {journal} {\bibinfo  {journal} {J. Appl. Phys.}\ }\textbf
  {\bibinfo {volume} {109}},\ \bibinfo {pages} {07A923} (\bibinfo {year}
  {2011})}\BibitemShut {NoStop}%
\bibitem [{\citenamefont {Mathieu}\ \emph {et~al.}(2011)\citenamefont
  {Mathieu}, \citenamefont {Ivanov}, \citenamefont {Tellgren},\ and\
  \citenamefont {Nordblad}}]{Mathieu11}%
  \BibitemOpen
  \bibfield  {author} {\bibinfo {author} {\bibfnamefont {R.}~\bibnamefont
  {Mathieu}}, \bibinfo {author} {\bibfnamefont {S.~A.}\ \bibnamefont {Ivanov}},
  \bibinfo {author} {\bibfnamefont {R.}~\bibnamefont {Tellgren}}, \ and\
  \bibinfo {author} {\bibfnamefont {P.}~\bibnamefont {Nordblad}},\ }\href@noop
  {} {\bibfield  {journal} {\bibinfo  {journal} {Phys. Rev. B}\ }\textbf
  {\bibinfo {volume} {83}},\ \bibinfo {pages} {174420} (\bibinfo {year}
  {2011})}\BibitemShut {NoStop}%
\bibitem [{\citenamefont {Tol\'{e}dano}\ and\ \citenamefont
  {Dmitriev}(1996)}]{Toledano}%
  \BibitemOpen
  \bibfield  {author} {\bibinfo {author} {\bibfnamefont {P.}~\bibnamefont
  {Tol\'{e}dano}}\ and\ \bibinfo {author} {\bibfnamefont {V.}~\bibnamefont
  {Dmitriev}},\ }\href@noop {} {\emph {\bibinfo {title} {Reconstructive phase
  transition: in crystals and quasicrystals}}}\ (\bibinfo  {publisher} {World
  Scientific Publishing, Singapore},\ \bibinfo {year} {1996})\BibitemShut
  {NoStop}%
\bibitem [{\citenamefont {Fisher}()}]{Fisher62}%
  \BibitemOpen
  \bibfield  {author} {\bibinfo {author} {\bibfnamefont {M.~E.}\ \bibnamefont
  {Fisher}},\ }\href@noop {} {\bibinfo  {journal} {Phil. Mag. ) \textbf{7},
  1731 (1962); P. Nordblad, L. Lundgren, E. Figueroa, and O. Beckman, J. Magn.
  Magn. Mater. \textbf{23}, 333 (1981)}\ }\BibitemShut {NoStop}%
\bibitem [{\citenamefont {Birss}(1966)}]{Birss}%
  \BibitemOpen
  \bibfield  {author} {\bibinfo {author} {\bibfnamefont {R.~R.}\ \bibnamefont
  {Birss}},\ }\href@noop {} {\emph {\bibinfo {title} {Symmetry and
  magnetism}}}\ (\bibinfo  {publisher} {North-Holland Publishing Company,
  Amsterdam},\ \bibinfo {year} {1966})\BibitemShut {NoStop}%
\end{thebibliography}

%

\end{document}